\documentclass[a4paper]{jpconf}
\usepackage{graphicx}
\usepackage{mathtools}
\usepackage{ifpdf}
\ifpdf        
  \usepackage{graphicx}
\else                          
  \usepackage[dvipdfmx]{graphicx}      
\fi
    
\def\lsim{\mathrel{\rlap{\lower4pt\hbox{\hskip1pt$\sim$}}
    \raise1pt\hbox{$<$}}}                
\def\gsim{\mathrel{\rlap{\lower4pt\hbox{\hskip1pt$\sim$}}
    \raise1pt\hbox{$>$}}}

\newcommand{\Slash}[1]{{\ooalign{\hfil \hspace*{-5pt}~#1\hfil\crcr\raise.167ex\hbox{/}}}}
\def\be{\begin{equation}}
\def\ee{\end{equation}}

\def\({\left(}
\def\){\right)}
\def\<{\langle}
\def\>{\rangle}
\newcommand{\non}{\nonumber \\ }
\newcommand{\matl}{\left( \begin{array}}
\newcommand{\matr}{\end{array} \right)}

\newcommand{\eq}[1]{Eq.~(\ref{#1})}

\def\beq#1\eeq{\begin{align}#1\end{align}}

\newcommand{\GEV}{\text{\,GeV} }

\newcommand{\real}{\textrm{Re}\,}
\newcommand{\imag}{\textrm{Im}\,}
\newcommand{\fig}[1]{Fig.~\ref{#1}}

\usepackage[symbol]{footmisc}

\begin{document}

\hfill {TTP16--048}

\title{Recent progress on $\boldsymbol{CP}$ violation in $\boldsymbol{K \to \pi \pi}$ decays in the SM and a supersymmetric solution
}

\author{T Kitahara$^{1,2,\ast}$\footnote[0]{$^{\ast}$ Speaker.},  U Nierste$^{1}$, and P Tremper$^{1}$
}
\address{
$^{1}${~Institute for Theoretical Particle Physics (TTP), Karlsruhe Institute of Technology, Engesserstra{\ss}e 7, D-76128 Karlsruhe, Germany}\\
 $^{2}${~Institute for Nuclear Physics (IKP), Karlsruhe Institute of
    Technology, Hermann-von-Helmholtz-Platz 1, D-76344
    Eggenstein-Leopoldshafen, Germany}
}

\ead{teppei.kitahara@kit.edu}

\begin{abstract}
  {
Using the recent first lattice results 
  of the  RBC-UKQCD collaboration  for $K \to \pi\pi$ decays, we perform a phenomenological analysis of $\epsilon_K^\prime/\epsilon_K$ and find a discrepancy between SM prediction and
experiments by $\sim 3\,\sigma$. We discuss an explanation by new physics. 
The well-understood  value of $\epsilon_K$, which quantifies indirect $CP$ violation,  however, typically prevents large new physics contributions to $\epsilon_K^\prime/\epsilon_K$.
 In this talk, we show a solution of the $\epsilon_K^\prime/\epsilon_K$ discrepancy in the Minimal Supersymmetric Standard Model with squark masses above 3 TeV without fine-tuning of $CP$ phases. 
 In this solution, the \emph{Trojan penguin} diagram gives large isospin-breaking contributions which enhance  $\epsilon_K^\prime$, 
while the contribution to $\epsilon_K$ is suppressed thanks to the Majorana nature of gluinos.
  }
\end{abstract}

\vspace{-1cm}
\subsubsection*{\hspace{-0.1cm}
\sl{KAON 2016 conference,\\
14-17 September 2016\\
University of Birmingham, United Kingdom}
}

\boldmath  
\section{Introduction to $\epsilon'_K/\epsilon_K$ discrepancy}\unboldmath  

\begin{figure}[t]
  \begin{center} 
    \includegraphics[width=0.5 \textwidth, 
    bb
     = 0 0 372 262]{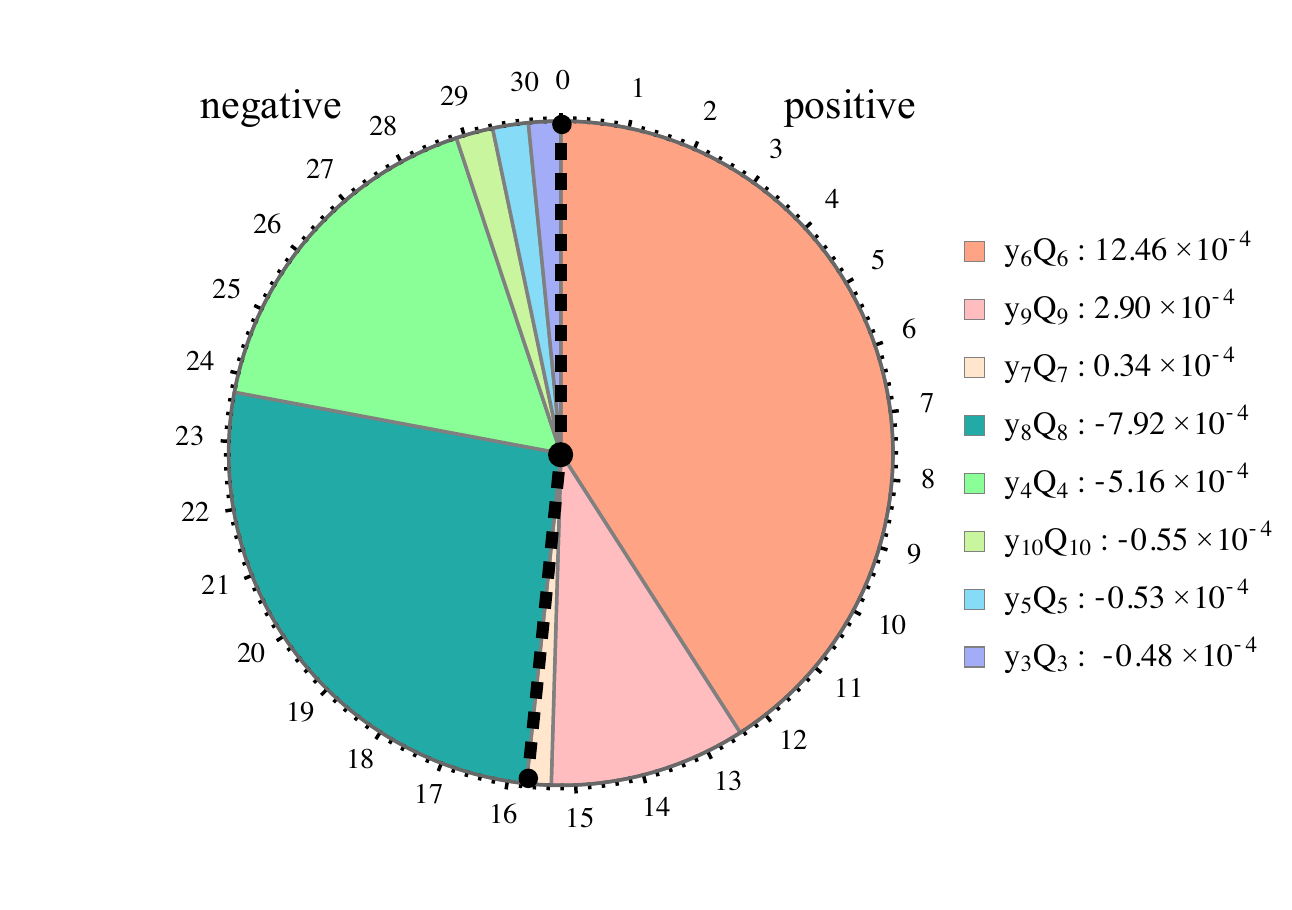}
  \caption{The composition of $  \epsilon_{K}'/\epsilon_{K}$ with respect to the operator basis. The right and left side of the dashed lines represent positive and negative contributions to $ \epsilon_{K}'/\epsilon_{K}$, respectively.  This figure is based on the result of Ref.~\cite{Kitahara:2016nld}. }
        \vspace{-0.4cm}
\label{fig:component}
\end{center}
\end{figure}

In $K \to \pi\pi$ decays, one distinguishes between two types of  charge-parity ($CP$)
violation: direct and indirect $CP$ violations which are parametrized by 
$\epsilon_{K}^{\prime}$ and $\epsilon_K$, respectively.
Both types of $CP$ violation have been quantified by many kaon experiments  precisely. 
While $\epsilon_K$ is a per mille effect in the Standard Model (SM), {$\epsilon_{K}^{\prime}$ is   smaller by another three orders of magnitude:
  } $\epsilon_{K}^{\prime} \sim \mathcal{O}(10^{-6})$.  This strong
suppression comes from the suppression of the isospin-$3/2$ amplitude w.r.t.~the isospin-$1/2$ amplitude ($\Delta I = 1/2$ rule)  and an
accidental cancellation of leading contributions in the SM.
In Fig.~\ref{fig:component}, the contributions of individual operators
to $ \epsilon_{K}'/\epsilon_{K}$ are shown. $Q_3$--$Q_6$  are called QCD penguin operators, while $Q_7$--$Q_{10}$  are called EW penguin operators.
The leading contributions come from $Q_6$ and $Q_8$, having opposite sign,  
and thus a cancellation emerges. 
Remarkably, this figure also shows that even if one includes sub-leading contributions, the cancellation  still exists with high precision.

\begin{figure}[t]
  \begin{center} 
    \includegraphics[width=0.8 \textwidth, 
    bb
     = 0 0 355 212]{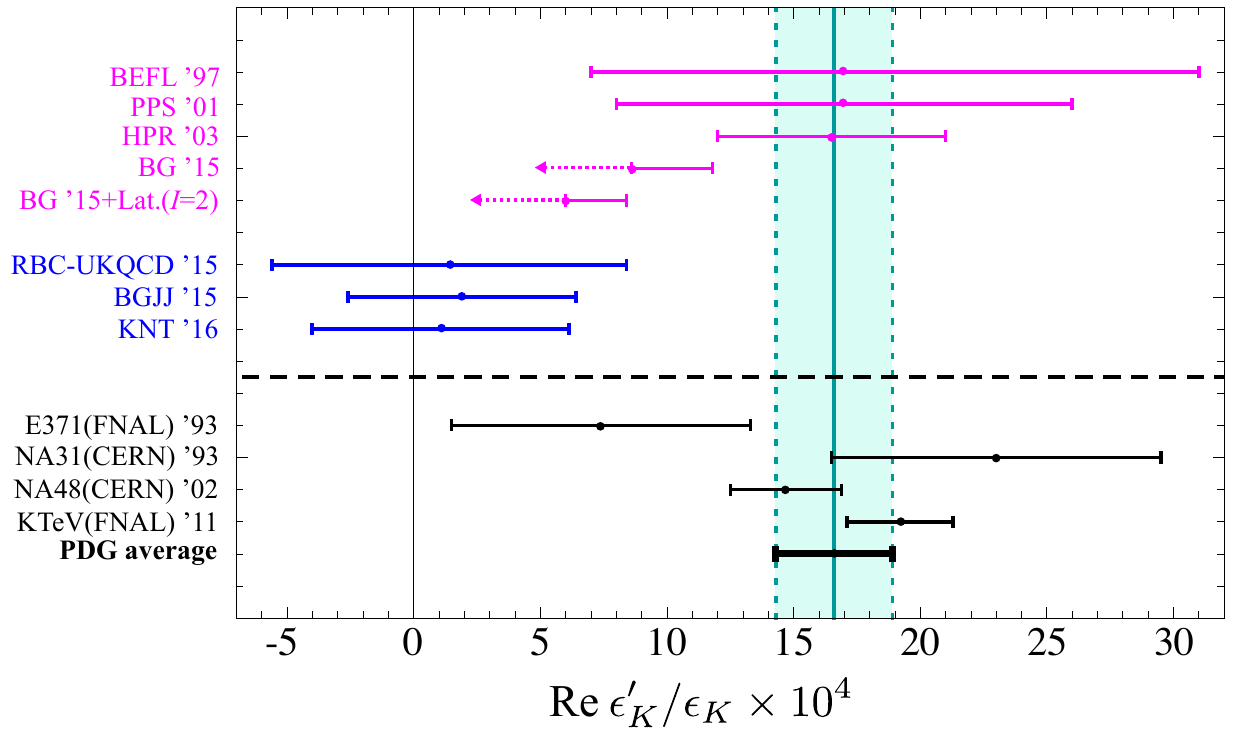}
    \caption{Compilation of representative SM predictions and the experimental values for Re~$\epsilon'_K / \epsilon_K$.  All error bars represent 1\,$\sigma$ range. The SM predictions are taken from Bertolini \emph{et al.} (BEFL '97) \cite{Bertolini:1997nf},  Pallante \emph{et al.} (PPS '01) \cite{Pallante:2001he}, Hambye \emph{et al.} (HPR '03) \cite{Hambye:2003cy}, Buras and G\'erard (BG '15) \cite{Buras:2015xba}, RBC-UKQCD lattice result \cite{Blum:2011ng}, Buras \emph{et al.} (BGJJ '15) \cite{Buras:2015yba}, and Kitahara \emph{et al.} (KNT '16)  \cite{Kitahara:2016nld}, where magenta bars are based on analytic approaches to hadronic matrix elements, while blue bars are based on lattice results.
    The black thick one is the world average of the experimental values \cite{Olive:2016xmw}.   }
          \vspace{-0.4cm}
\label{fig:status}
\end{center}
\end{figure}

The compilation of representative SM predictions and the experimental values for Re~$\epsilon'_K / \epsilon_K$ is given in Fig.~\ref{fig:status}.
The SM predictions (colored bars) are taken from:
 Bertolini \emph{et al.} (BEFL '97) \cite{Bertolini:1997nf},  Pallante \emph{et al.} (PPS '01) \cite{Pallante:2001he}, Hambye \emph{et al.} (HPR '03) \cite{Hambye:2003cy}, Buras and G\'erard (BG '15) \cite{Buras:2015xba} with lattice result for $I=2$ (BG '15$+$Lat.), RBC-UKQCD lattice result \cite{Blum:2011ng}, Buras \emph{et al.} (BGJJ '15) \cite{Buras:2015yba}, and Kitahara \emph{et al.} (KNT '16)  \cite{Kitahara:2016nld}.
The experimental values (black bars) are taken from: 
 E371 \cite{Gibbons:1993zq}, NA31 \cite{Barr:1993rx}, NA48 \cite{Fanti:1999nm} and KTeV \cite{AlaviHarati:1999xp} collaborations,
and the black thick one is the world average of the experimental values \cite{Olive:2016xmw},
\beq%
   \textrm{Re}\left( \frac{\epsilon_K^\prime}{ \epsilon_K} \right)=
\left(16.6 \pm 2.3 \right) \times 10^{-4}~(\textrm{PDG~average}).
  \label{PDG}
  \eeq %

In order to predict $\epsilon_K^\prime$ in the SM, one has to calculate the hadronic matrix elements of four-quark operators with nonperturbative methods.
The magenta bars  in Fig.~\ref{fig:status}  have utilized analytic approaches  to calculating  them: 
 chiral quark model (BEFL '97),  chiral perturbation theory (PPS '01) with  minimal hadronic approximation (HPR '03), and the dual   QCD approach (BG '15). Note that the dual QCD approach predicts an upper bound on $\epsilon'_K / \epsilon_K$.
 On the other hand, a determination of all hadronic matrix elements from lattice QCD has  been obtained only recently by the RBC-UKQCD collaboration \cite{Blum:2011ng}, and 
 the blue bars are based on the lattice result:
\beq%
\frac{\epsilon_K^\prime}{\epsilon_K}=
\begin{cases}
  \left(1.9 \pm 4.5  \right) \times 10^{-4}&(\textrm{BGJJ~'15}),\\
 {\left(1.06 \pm  5.07 \right)}\times 10^{-4}&(\textrm{KNT~'16}).
  \end{cases}
  \label{discrepancy}
  \eeq %
  These results are obtained by next-to-leading order (NLO) calculations exploiting 
    $CP$-conserving data to reduce hadronic uncertainties and include isospin-violating contributions \cite{Cirigliano:2003nn} which are not included in the lattice result.  
    Furthermore, the latter result includes an additional $\mathcal{O}(\alpha_{EM}^2/\alpha^2_s)$ correction, which appears only in this order, and also
  utilizes a new analytic solution of the renormalization group (RG) equation which avoids the  problem of singularities in the NLO terms.
   The two
    numbers in~\eq{discrepancy} disagree with the experimental
value in \eq{PDG}  by $2.9\,\sigma$ \cite{Buras:2015yba} and $2.8\,\sigma$
    {\cite{Kitahara:2016nld}}, respectively. 
   The  uncertainties are  dominated by the lattice statistical and systematic uncertainties for the $I = 0$ amplitude.
   Therefore, in the near future, the increasing precision of lattice calculations will further sharpen  the SM
  predictions in \eq{discrepancy} and  answer the question about  new physics (NP) in $\epsilon_K^\prime/\epsilon_K$.

The main difference between each result of analytic approaches and the lattice result  is in the hadronic parameter  
$B_6^{(1/2)}$, which controls the largest  positive  contribution to $\epsilon'_K/\epsilon_K$, the $y_6 Q_6$ component in the Fig.~\ref{fig:component}.
In chiral perturbation  theory, typically large values are obtained: $B_6^{(1/2)} \sim 1.6 $ (BEFL '97), $\sim1.6$ (PPS '01), and $\sim3$ (HPR '03, \cite{Buras:2015xba}).
On the other hand, the dual QCD approach predicts a smaller number, $B_6^{(1/2)} \leq  B_8^{(3/2)} \sim  0.8$ (BG '15).
The lattice result is consistent with the latter result: $B_6^{(1/2)}  = 0.56 \pm 0.20$ \cite{Blum:2011ng, Kitahara:2016nld}.
 Note that the lattice calculation  \emph{includes}\ final-state interaction of the two pions in accordance with Ref.~\cite{Lellouch:2000pv}.

We also should comment on the $\Delta I = 1/2$ rule, which  denotes the largeness  of the ratio of the $CP$-conserving amplitudes, 
$(\textrm{Re} A_0 / \textrm{Re} A_2)_{\textrm{exp.}}  = 22.45 \pm  0.05$.
Although none of the analytic approaches can explain such a large value,  the first lattice calculation has found a consistent value within $1\,\sigma$, 
$(\textrm{Re} A_0 / \textrm{Re} A_2)_{\textrm{Lat.}}  = 31.0 \pm  11.1$ \cite{Blum:2011ng, Buras:2016fys}.

\boldmath  
\section{$\epsilon_K$ in the MSSM}\unboldmath

An explanation of the puzzle between \eq{PDG} and  \eq{discrepancy} by physics beyond the  SM requires a NP contribution which is seemingly even larger than the SM contribution. However, it is known that once constraints from the corresponding $|\Delta S|=2$ transition are taken into account, one expects that NP effects in a $|\Delta S|=1$ four-quark process are highly suppressed.
  To explain the NP hierarchy in $|\Delta S|=1$ vs $|\Delta S|=2$ transitions, we specify to   $\epsilon_K^\prime$ and $\epsilon_K$: The SM contributions are governed by the combination
  $ \tau = -  V_{td}V_{ts}^* /(V_{ud}V_{us}^*) \sim (1.5 - i 0.6)\times 10^{-3} $
with
  $\epsilon_K^{\prime\,\rm SM} \propto \imag \tau/M_W^2$ and $\epsilon_K^{\rm SM} \propto \imag \tau^2/M_W^2$. 
   If the NP contribution enters through  the $\Delta S=1$ parameter $\delta$ and is
  mediated by heavy particles of mass $M$, one obtains
  $\epsilon_K^{\prime\, \mathrm{NP}} \propto \imag \delta/M^2$, $\epsilon_K^{\mathrm{NP}} \propto \imag \delta^2/M^2$, and therefore
  {the experimental constraint} {$|\epsilon_K^{\mathrm{NP}}| \leq
    |\epsilon_K^{\rm SM}|$ leads to}%
\beq%
  {\left| {\frac{\epsilon^{\prime\,
          \mathrm{NP}}_K}{\epsilon^{\prime\, \mathrm{SM}}_K}
      } \right|} \leq  
  \frac{\left|\epsilon_K^{\prime\,
         \mathrm{NP}}/\epsilon_K^{\prime\,\rm SM} \right|}{
     \left|\epsilon_K^{\mathrm{NP}}/\epsilon_K^{\rm SM} \right|}
  = {\cal O}\left( \frac{\real \tau}{\real \delta} \right).
\label{eq:sens}
\eeq%
If NP enters through a loop with particles of mass $M\gsim 1$ TeV,   the NP effects can be relevant only for $|\delta|\gg
|\tau|$, and  thus \eq{eq:sens} seemingly forbids detectable NP contributions to $\epsilon_K^\prime$.

In the  Minimal Supersymmetric Standard Model (MSSM), there is a  bypass to  \eq{eq:sens}.
 The Majorana nature of the gluino leads to a suppression of gluino-squark box contributions to $\epsilon_K$.
This is so, because there are two such diagrams (crossed and  uncrossed boxes) with opposite signs. 
If the gluino mass $m_{\tilde g}$ equals roughly 1.5 times the average down squark 
mass $M_S$,  both contributions to $\epsilon_K^{\rm SUSY}$ cancel \cite{Crivellin:2010ys}. 
For $m_{\tilde g}>1.5 M_S $, the gluino-box contribution approximately behaves as $[m_{\tilde g}^2-(1.5 M_S)^2]/m_{\tilde g}^4$,  and the  $1/m_{\tilde g}^2$ decoupling sets in. 
Note that this suppression appears only when a hierarchy $\Delta_{Q,12} \gg \Delta_{\bar{D},12}$ of  $\Delta_{Q,12} \ll  \Delta_{\bar{D},12}$ is satisfied, where the following notation is used for the squark mass matrices: $ M^2_{X, ij} = m^2_{X} \left( \delta_{ij} + \Delta_{X, ij} \right), $
with $X = Q,~\bar{U}$,~or~$\bar{D}$.

\boldmath  
\section{$\epsilon_K^\prime/ \epsilon_K$ in the MSSM}\unboldmath

\begin{figure}[tb] 
 \begin{center} 
    \includegraphics[width=0.50 \textwidth, 
    bb 
    = 0 0 1753 544]{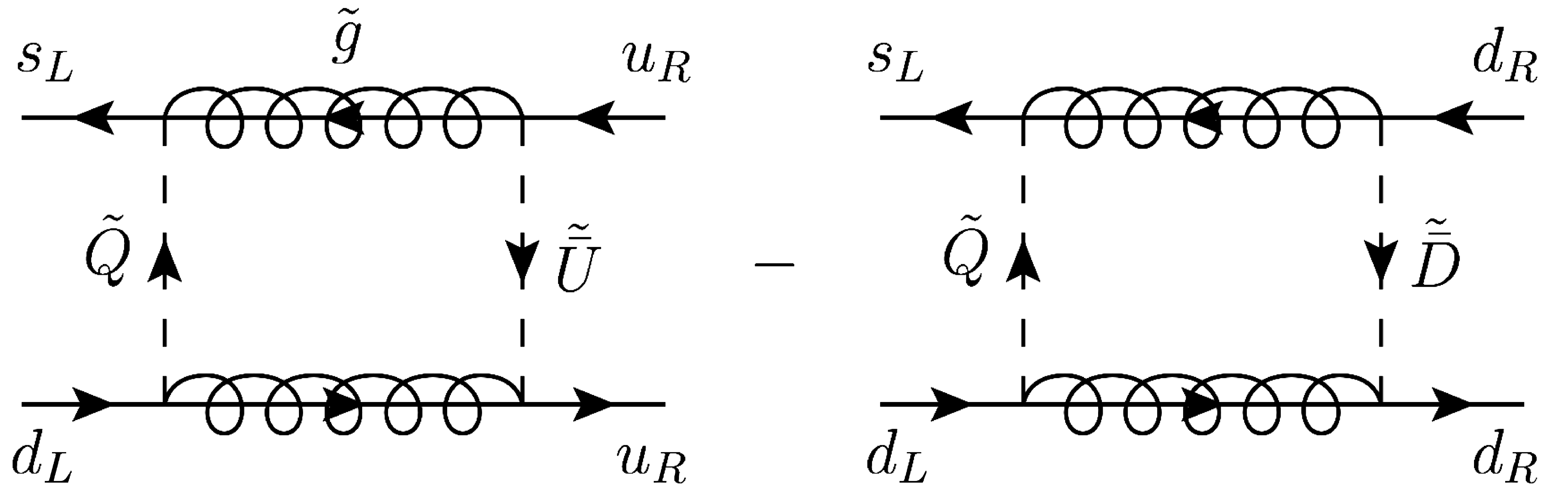}
    \caption{{\it Trojan
       penguin}\ contributions to $\imag A_2$ for
      $m_{\bar{U}}\neq m_{\bar{D}}$.}
      \vspace{-0.6cm}
      \label{fig:trojan}
\end{center} 
\end{figure} 
The master equation for $\epsilon_K^\prime$ is given by \cite{Buras:2015yba}
\beq%
\frac{\epsilon_K^\prime}{\epsilon_K} = \frac{\omega_{+}}{\sqrt{2}
  {|}\epsilon_K^{\textrm{exp}}{|} \textrm{Re} A_0^{\textrm{exp}} } \left\{
  \frac{\textrm{Im} A_2 }{\omega_{+}} - \left( 1-
    \hat{\Omega}_{\textrm{eff}} \right) \textrm{Im} A_0 \right\},
\label{eq:mas}
\eeq%
with $\hat{\Omega}_{\textrm{eff}} = (14.8\pm
8.0)\times 10^{-2}$,  {the measured
  $|\epsilon_K^{\rm exp}|$,} $\omega_{+}= (4.53 \pm 0.02)\times10^{-2}$,  and the amplitudes $A_I = \langle (\pi \pi)_I |
\mathcal{H}^{\left|\Delta S\right| = 1} | K^0 \rangle$ involving the
effective $|\Delta S|=1$ {Hamiltonian} $ \mathcal{H}^{\left|\Delta
    S\right|}$. $I=0,2$ represents the strong isospin of the final two-pion
state.

The MSSM contributions to $\epsilon_K^\prime/\epsilon_K$ have been widely studied in the past.
However, the supersymmetry-breaking scale $M_S$  was considered in the  ballpark of  the electroweak scale, so that
the suppression mechanism inferred from \eq{eq:sens} is avoided. 
The low-energy Hamiltonian in the case of small left-right squark mixing reads
\beq
\mathcal{H}_{\textrm{eff, SUSY}}^{\left|\Delta S \right| = 1} =
\frac{G_F}{\sqrt{2}}\sum_q \left[ \sum_{i=1}^{2} c_i^{q} (\mu) Q_i^q
  (\mu)   +\sum_{i=1}^4  [
    c_i^{\prime q} (\mu) Q_i^{\prime q}(\mu) + \tilde{c}_i^{\prime q}
    (\mu) \tilde{Q}_i^{\prime q}(\mu) ] \right] + \textrm{H.c.},
\label{eq:hsusy}
\eeq
where $G_F$ is the Fermi constant and
\begin{eqnarray}
Q^{ q}_1 &= \left( \bar{s}_{\alpha} q_{\beta} \right)_{{}_{V-A}}\left( \bar{q}_{\beta} d_{\alpha} \right)_{{}_{V-A}},~~~
Q^{ q}_2 = \left( \bar{s}  q \right)_{{}_{V-A}} \left( \bar{q}  d \right)_{{}_{V-A}},\non
Q^{\prime q}_1 &= \left( \bar{s}  d \right)_{{}_{V-A}} \left( \bar{q}  q \right)_{{}_{V+A}},~~~~
Q^{\prime q}_2 = \left( \bar{s}_{\alpha}  d_{\beta} \right)_{{}_{V-A}} \left( \bar{q}_{\beta}  q_{\alpha} \right)_{{}_{V+A}}, \non
Q^{\prime q}_3 &= \left( \bar{s}  d \right)_{{}_{V-A}} \left( \bar{q}  q \right)_{{}_{V-A}},~~~~
Q^{\prime q}_4 = \left( \bar{s}_{\alpha}  d_{\beta} \right)_{{}_{V-A}}  \left( \bar{q}_{\beta}  q_{\alpha} \right)_{{}_{V-A}}.
\end{eqnarray}
Here $( \bar{s} d)_{V-A} ( \bar{q} q )_{V\pm A}= [\bar{s}\gamma_{\mu}(1-\gamma_5)d][ \bar{q} \gamma^{\mu}(1\pm \gamma_5) q]$, $\alpha$ and $\beta$ represent color indices, and opposite-chirality operators $ \tilde{Q}_i^{\prime q}$ are given by interchanging $V-A \leftrightarrow V+A$. 

\begin{figure}[t]
  \begin{center} 
    \includegraphics[width=0.50 \textwidth, 
    bb
     = 0 0 360 362]{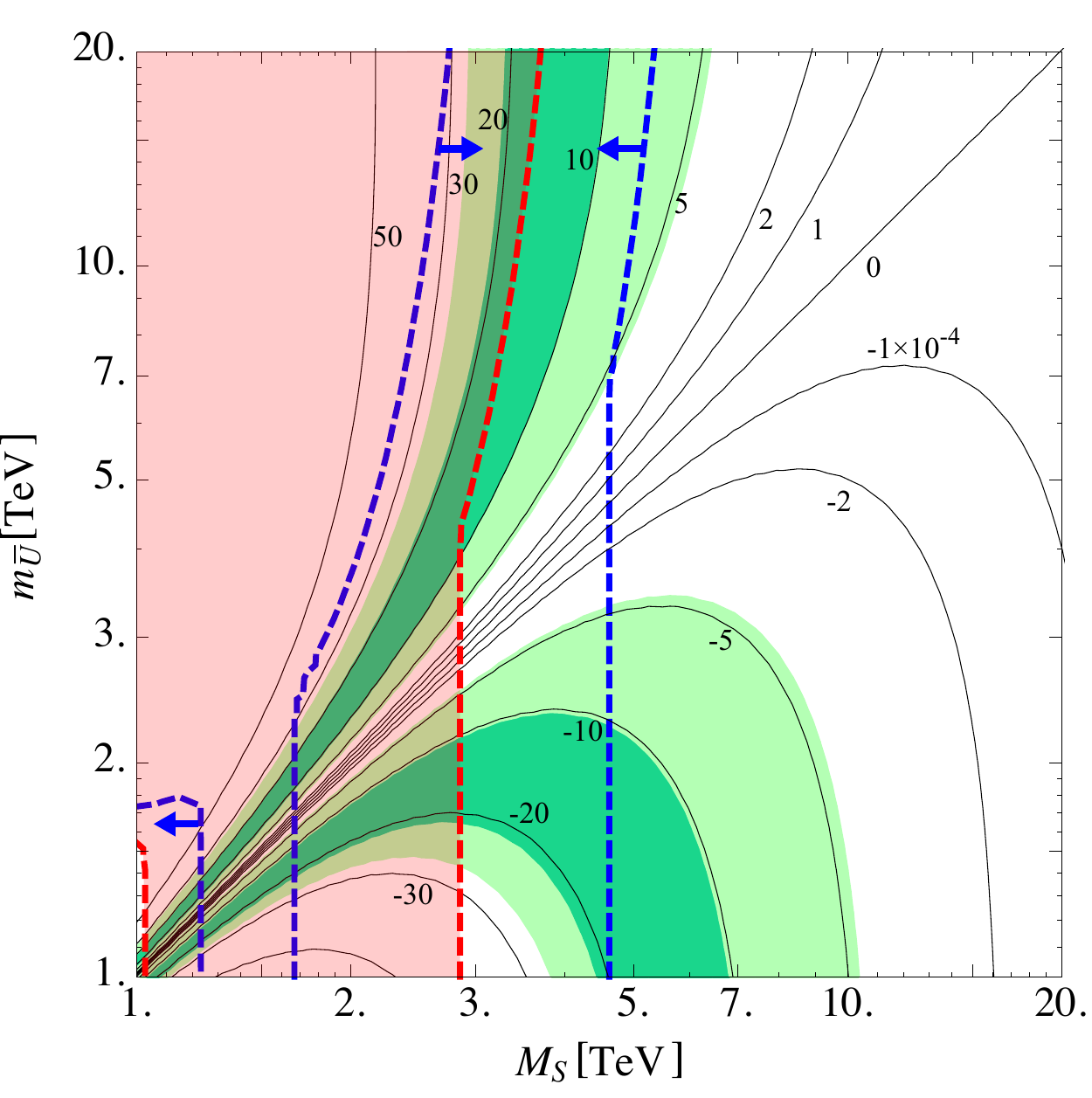}   
    \caption{The black contour represents the supersymmetric contributions to $\epsilon_K^\prime/\epsilon_K$ in units of $10^{-4}$.  
      The $\epsilon_K^\prime/\epsilon_K$ discrepancy is resolved at
     $1\,\sigma$\,($2\,\sigma$) in the dark (light) green region.  
     {The} red  shaded  region (region between the blue dashed lines) is excluded (preferred) by $\epsilon_K$ with inclusive (exclusive) $|V_{cb}|$ at
     95\,\% C.L.  
     }
     \vspace{-.6cm}
\label{fig:contour}
\end{center}
\end{figure}
  
In our analysis \cite{Kitahara:2016otd} we found that  the dominant supersymmetric contribution comes from a certain gluino box diagrams$^{\#1}$\footnote[0]{$^{\#1}$ The other supersymmetric solution focusing the chargino $Z$-penguin contribution has been studied in Ref.~\cite{Tanimoto:2016yfy}.
  }, 
  called a {\it Trojan penguin}\ in Ref.~\cite{Kagan:1999iq}, which are shown in  \fig{fig:trojan}.
  It contributes to $\imag A_2$ when $m_{\bar{U}}\neq m_{\bar{D}}$. Because these contributions are  governed by the strong interaction and there is  an enhancement factor $1/\omega_+ = 22.1$ for the $\textrm{Im} A_2$ term in \eqref{eq:mas}, they easily become the largest contribution to $\epsilon'_K / \epsilon_K$
\cite{Kagan:1999iq}. 
In order to obtain the desired large effect in $\epsilon_K^\prime$, one needs a contribution to the operators $Q_{1,2}^\prime$ with $(V-A)\times(V+A)$ Dirac structure, whose matrix elements are chirally enhanced by a 
factor $(m_K/m_s)^2$. 
Hence, the flavor mixing has to be in the left-handed squark mass matrix.  The opposite situation with right-handed flavor mixing and $\tilde u_L$-$\tilde d_L$ mass splitting is not possible because of the SU(2)$_L$ invariance.

{For the calculation of supersymmetric contributions to $\epsilon_K^\prime/ \epsilon_K$, one has to use the RG equations to evolve the Wilson coefficients calculated at the high scale $M_S$ down to the hadronic scale $\mu_h = {\cal O} (1\,\GEV)$ at which the hadronic matrix elements are
calculated \cite{Blum:2011ng, Kitahara:2016nld}.}   
To use the well-known NLO $10\times10$ anomalous dimensions for the SM four-fermion operator basis \cite{nlo}, 
we switch from \eq{eq:hsusy} to%
\beq%
\mathcal{H}_{\textrm{eff, SUSY}}^{\left| \Delta S \right| =1 } =
\frac{G_F}{\sqrt{2}} \sum_{i=1}^{10} [ C_{i} (\mu) Q_i (\mu) +
  \tilde{C}_{i} (\mu) \tilde{Q}_i (\mu) ] + \textrm{H.c.}, %
\eeq%
where $Q_{1,\dots,10}$ are given in Ref.~\cite{nlo} 
and 
\beq
C_{1,2} (\mu) = c_{1,2}^u (\mu),&\qquad \tilde{C}_{1,2}(\mu)= 0, \non
C_{3,4,5,6} (\mu) = \frac{1}{3} [ c_{3,4,1,2}^{' u}(\mu) + 2
  c_{3,4,1,2}^{' d} (\mu) ], &\qquad C_{7,8,9,10} (\mu) =
\frac{2}{3}[ c_{1,2,3,4}^{'u}(\mu) - c_{1,2,3,4}^{'d} (\mu)
],  
\eeq
and the coefficients $\tilde{C}_{3,\dots10}$ for the opposite-chirality operators can be {obtained from $C_{3,\dots10}$ by
  replacing} $c^{'q}_i \to \tilde{c}^{'q}_i$.  
  Note that the contribution of Fig.~\ref{fig:trojan} is collected into  the coefficients $C_{7,8} $.
{For the RG evolution of the coefficients, we use the new analytic solution  of the RG equations  discussed in {Ref.~\cite{Kitahara:2016nld}}.

Our main result is given in Fig.~\ref{fig:contour}, where the portion of the squark mass plane which simultaneously explains
$\epsilon_K^\prime/\epsilon_K$ and $\epsilon_K$ is shown.    
As input, we take the grand-unified theory relation for gaugino masses, $\alpha_s \left( M_Z \right) = 0.1185$, $m_{\tilde{g}}/M_S = 1.5 $ for the suppressed $\epsilon_K$, and $m_{Q} = m_{\bar{D}} = \mu_{\rm SUSY}=M_S$ with varying $m_{\bar U}$}.
Furthermore, the trilinear supersymmetry-breaking matrices $A_q$ are set to zero, $\tan \beta =10$, and the only nonzero off-diagonal elements of the squark mass  matrices are $\Delta_{Q, 12,13,23} = 0.1 \exp(- i \pi/4)$  {for the left-handed squark sectors}.
We have calculated all relevant one-loop contributions to the coefficients in \eq{eq:hsusy} in the squark mass eigenbasis.
The $\epsilon_K^\prime/\epsilon_K$ discrepancy can be resolved at $1\,\sigma$\,($2\,\sigma$) in the dark (light) green region.  
The red region is already excluded by the measurement of $\epsilon_K$ at 95\,\% C.L. in combination with the inclusive
$V_{cb}$. On the other hand, the region between the blue dashed lines can explain the $\epsilon_K$ discrepancy at 95\,\% C.L.\ for the exclusive value of $|V_{cb}|$.  
The area in Fig.~\ref{fig:contour} labeled with negative values of $\epsilon^{\prime}_K /\epsilon_K $  becomes feasible by adding $\pi$ to the phase of $\Delta_{Q,ij}$, which flips the sign of $\epsilon^{\prime}_K $  while keeping $\epsilon_K$ unchanged.

\section{Conclusions}

In this talk, we have discussed the supersymmetric contributions to $ \epsilon_K^\prime$,  and it is found that the large contributions required to solve the discrepancy between \eq{PDG} and  \eq{discrepancy} can be obtained in the multi-TeV squark mass range thanks to the Trojan penguin diagrams. 
Using a relatively heavy gluino, the severe constraint from $ \epsilon_K$ can be fulfilled without fine-tuning.

\vspace{-.2cm}

\section*{Acknowledgments}{
The work of UN is supported by BMBF under
grant no.~05H15VKKB1.  PT acknowledges support from the DFG-funded
doctoral school {\it KSETA}.
}


\section*{References}
\begin{thebibliography}{99}
\expandafter\ifx\csname url\endcsname\relax
  \def\url#1{{\tt #1}}\fi
\expandafter\ifx\csname urlprefix\endcsname\relax\def\urlprefix{URL }\fi
\providecommand{\eprint}[2][]{\url{#2}}


\bibitem{Bertolini:1997nf} 
  Bertolini S, Eeg J~O, Fabbrichesi M  and Lashin E I  1998
  Nucl.\ Phys.\ B {\bf 514}, 93 
  (\textit{Preprint} \eprint{hep-ph/9706260})

\bibitem{Pallante:2001he} 
Pallante   E and Pich A 2001 
  Nucl.\ Phys.\ B {\bf 592}, 294 
    (\textit{Preprint} \eprint{hep-ph/0007208}),  
Pallante  E, Pich A  and Scimemi I 2001
  Nucl.\ Phys.\ B {\bf 617}, 441 
  (\textit{Preprint} \eprint{hep-ph/0105011})
  
\bibitem{Hambye:2003cy} 
Hambye   T, Peris S and de Rafael  E 2003
  JHEP {\bf 0305}, 027 
    (\textit{Preprint} \eprint{hep-ph/0305104})
  
\bibitem{Buras:2015xba} 
  Buras A J and G\'erard J M 2015 
  JHEP {\bf 1512}, 008
   (\textit{Preprint} \eprint{1507.06326 [hep-ph]})
  
\bibitem{Blum:2011ng} 
Blum   T {\it et al} 2012
     {Phys.\ Rev.\ Lett.\  {\bf 108}, 141601 
        (\textit{Preprint} \eprint{1111.1699 [hep-lat]})},
 Blum T {\it et al} 2012
{Phys.\ Rev.\ D {\bf 86}, 074513 
      (\textit{Preprint} \eprint{1206.5142 [hep-lat]})},
  Blum T {\it et al} 2015
 {Phys.\ Rev.\ D {\bf 91}, no. 7, 074502 
      (\textit{Preprint} \eprint{1502.00263 [hep-lat]})},
Bai   Z {\it et al} [RBC and UKQCD Collaborations]  2015
{Phys.\ Rev.\ Lett.\  {\bf 115}, no. 21, 212001
      (\textit{Preprint} \eprint{1505.07863 [hep-lat]})}
  
\bibitem{Buras:2015yba} 
Buras   A J, Gorbahn M, J\"ager S  and Jamin M 2015
{JHEP {\bf 1511}, 202 
 (\textit{Preprint} \eprint{1507.06345 [hep-ph]})}
  
\bibitem{Kitahara:2016nld} 
Kitahara   T, Nierste U and Tremper P 
   (\textit{Preprint} \eprint{1607.06727 [hep-ph]})
  
\bibitem{Gibbons:1993zq} 
Gibbons    L K {\it et al} 1993
{Phys.\ Rev.\ Lett.\  {\bf 70}, 1203 }
  
\bibitem{Barr:1993rx} 
  Barr G D {\it et al} [NA31 Collaboration] 1993
     {Phys.\ Lett.\ B {\bf 317}, 233}
    
\bibitem{Fanti:1999nm} 
Fanti    V {\it et al} [NA48 Collaboration] 1999
  {Phys.\ Lett.\ B {\bf 465}, 335 
   (\textit{Preprint} \eprint{hep-ex/9909022})},
Batley   J R  {\it et al} [NA48 Collaboration] 2002
   {Phys.\ Lett.\ B {\bf 544}, 97 
     (\textit{Preprint} \eprint{hep-ex/0208009})}
  
\bibitem{AlaviHarati:1999xp} 
Alavi-Harati A  {\it et al} [KTeV Collaboration] 1999
   {Phys.\ Rev.\ Lett.\  {\bf 83}, 22 
        (\textit{Preprint} \eprint{hep-ex/9905060})},
Abouzaid  E {\it et al} [KTeV Collaboration] 2011
{Phys.\ Rev.\ D {\bf 83}, 092001 
  (\textit{Preprint} \eprint{1011.0127 [hep-ex]})}
  
  
\bibitem{Olive:2016xmw} 
Patrignani  C 2016
  Chin.\ Phys.\ C {\bf 40}, no. 10, 100001 
  
\bibitem{Cirigliano:2003nn} 
Cirigliano V, Pich A, Ecker G and Neufeld H 2003
 {Phys.\ Rev.\ Lett.\  {\bf 91}, 162001 
  (\textit{Preprint} \eprint{hep-ph/0307030}) }
  
    
  \bibitem{Lellouch:2000pv} 
Lellouch  L and Luscher M 2001 
{Commun.\ Math.\ Phys.\  {\bf 219},  31
    (\textit{Preprint} \eprint{hep-lat/0003023})}
  
\bibitem{Buras:2016fys} 
Buras   A J and G\'erard J M
    (\textit{Preprint} \eprint{1603.05686 [hep-ph]})
  
  
\bibitem{Crivellin:2010ys} 
Crivellin  A  and Davidkov M 2010
  Phys.\ Rev.\ D {\bf 81}, 095004 
      (\textit{Preprint} \eprint{1002.2653 [hep-ph]})
      
      
\bibitem{Kitahara:2016otd} 
Kitahara   T, Nierste U and Tremper P 2016
  Phys.\ Rev.\ Lett.\  {\bf 117}, no. 9, 091802 
    (\textit{Preprint} \eprint{1604.07400 [hep-ph]})
  
\bibitem{Tanimoto:2016yfy} 
Tanimoto   M and Yamamoto K 
      (\textit{Preprint} \eprint{1603.07960 [hep-ph]}),
Endo   M, Mishima S, Ueda  D and Yamamoto K   2016
  Phys.\ Lett.\ B {\bf 762}, 493
        (\textit{Preprint} \eprint{1608.01444 [hep-ph]})
  
  
\bibitem{Kagan:1999iq} 
Kagan   A L and Neubert  M 1999
  Phys.\ Rev.\ Lett.\  {\bf 83}, 4929 
        (\textit{Preprint} \eprint{hep-ph/9908404}),
Grossman   Y, Neubert M and Kagan A L 1999
  JHEP {\bf 9910}, 029 
          (\textit{Preprint} \eprint{hep-ph/9909297})

  
  
  \bibitem{nlo} 
Buras A J, Jamin M, Lautenbacher M E and Weisz P H 1993
 {Nucl.\ Phys.\ B {\bf 400}, 37 
           (\textit{Preprint} \eprint{hep-ph/9211304}),}
Buras   A J, Jamin M and Lautenbacher M E  1993
{Nucl.\ Phys.\ B {\bf 400}, 75
       (\textit{Preprint} \eprint{hep-ph/9211321}),}
 Ciuchini  M, Franco E, Martinelli G and Reina L 1994 
 {Nucl.\ Phys.\ B {\bf 415}, 403 
      (\textit{Preprint} \eprint{hep-ph/9304257})}

  
  
  
\end{thebibliography}

\end{document}